\documentclass[12pt]{article}
\pdfoutput=1
\usepackage{authblk}
\usepackage{hyperref} 
\usepackage{hanging}

\title{{\bf\large{Three Aspects of Typicality in Multiverse Cosmology}}}

\author{Feraz Azhar\thanks{Email: fa232@cam.ac.uk}}
\affil{Department of History and Philosophy of Science, University of Cambridge, Free School Lane, Cambridge CB2 3RH, United Kingdom, and \\ Program in Science, Technology, and Society, Massachusetts Institute of Technology, Cambridge, MA 02139, United States of America}

\date{{\small\today}}

\begin{document}

\maketitle

\begin{abstract} 
Extracting predictions from cosmological theories that describe a multiverse, for what we are likely to observe in our domain, is crucial to establishing the validity of these theories. One way to extract such predictions is from theory-generated probability distributions that allow for selection effects---generally expressed in terms of assumptions about anthropic conditionalization and how typical we are. In this paper, I urge three lessons about typicality in multiverse settings. (i) Because it is difficult to characterize our observational situation in the multiverse, we cannot assume that we are typical (as in the `principle of mediocrity'): nor can we ignore the issue of typicality, for it has a measurable impact on predictions for our observations. (ii) There are spectra of assumptions about both conditionalization and typicality, which lead to coincident predictions for our observations, leading to problems of confirmation in multiverse cosmology. And moreover, (iii) when one has the freedom to consider competing theories of the multiverse, the assumption of typicality may not lead to the highest likelihoods for our observations. These three entwined aspects of typicality imply that positive assertions about our typicality, such as the `principle of mediocrity', are more questionable than has been recently claimed. 

\end{abstract}

\section{{\bf\large{Introduction}}}\label{SEC:Introduction}

A startling prediction of a range of modern cosmological theories is that there exist domains outside our observable horizon, where the fundamental constants of nature, and perhaps the effective laws of physics more generally, vary. These multi-domain universes (henceforth `multiverses') can be described by inflationary theories (Steinhardt 1983, Vilenkin 1983, Linde 1983, 1986a, 1986b), and have also attracted attention as a result of the discovery of the string theory landscape (Bousso and Polchinski 2000, Kachru et al.~2003, Susskind 2003). 
	
In this context, a pressing question is: how can we ascertain the existence of such a multiverse? Short of directly observing such a structure, one test is via a comparison between theory-generated probability distributions for observables, and observations that we make in our domain. Defining probability distributions in multiverse scenarios is, however, beset with difficulties (Aguirre 2007). One must select a measure, and deal with infinities that are often associated with such measures; and even if one makes headway on such issues, the presumed parsimony of description of any fundamental theory that describes a multiverse will most likely render probabilities for our observations very small (Hartle 2007). To facilitate the required comparison between theory-generated probability distributions and our observations, it has been argued that anthropic conditionalization is needed. But it is problematic how to achieve such conditionalization, since it is unclear who or what we should be conditionalizing on, as well as which physical parameters are needed to describe the object of the conditionalization.
	
Even if one manages to address these problems in a plausible way, a third stumbling block remains, which will constitute the focus of this paper. Namely, a suitable measure and conditionalization scheme might make our observations more likely: but the question remains how likely or \emph{typical} should they be, before we can consider them to have  provided support for the theory under consideration? 
	
One means to address this question is through the `principle of mediocrity' (so-named, and advocated, by Vilenkin in various works, e.g., Vilenkin 1995), which in more current terminology, claims that we should assume that we are typical of any reference class to which we believe we belong. Thus a given cosmological theory, measure, and suitable conditionalization scheme, which in combination, effectively define this reference class, will give rise to a probability distribution whose typical values constitute its predictions for what we may observe. The argument then goes that if we indeed observe these typical values, then our observations provide support for the conjunction of the theory, measure, and conditionalization scheme being considered. 
	
In this paper, I systematically present, for a philosophical audience, recent results that explore assumptions about typicality in three complementary settings (Azhar 2014, 2015, 2016). First, in section~\ref{SEC:TypMatters}, following Azhar (2014), I argue that under what is called top-down conditionalization [following Aguirre and Tegmark (2005)], namely, when we conditionalize our distributions by demanding consistency with (all relevant) experimental evidence, we cannot simply assume typicality [as argued by Garriga and Vilenkin (2008)], nor can we effectively ignore typicality (Weinstein 2006). I maintain, via a concrete example related to dark matter, that typicality dramatically affects top-down predictions, exemplifying the sense in which errors in reasoning about typicality translate to errors in the assessment of predictive power. I thereby advocate a line of thinking promoted by Srednicki and Hartle (2010): who argue for the inclusion of `xerographic distributions' in the computation of probability distributions for observables, where these xerographic distributions encode a variety of assumptions about typicality. One thus effectively tests a `framework' (in their terminology); namely, the conjunction of four items: a theory, a measure, a conditionalization scheme, and a typicality assumption as given by a xerographic distribution.

Second, in section~\ref{SEC:Confirmation}, following Azhar (2016), I argue that assumptions about typicality matter in the sense that different frameworks can give rise to the same prediction for what we are likely to observe in our domain. Thus there is a significant and problematic under-determination of multiverse frameworks by data. 
	
Third, in section~\ref{SEC:testingTypicality}, following Azhar (2015), I report on a test of the principle of mediocrity, which uses xerographic distributions in a multiverse model that generalizes a cosmological model of Hartle and Srednicki (2007, 2010). I argue that for a \emph{fixed} theory, the assumption of typicality gives rise to the highest likelihoods for our data. If, however, one allows the underlying theory and the assumption of typicality to \emph{vary}, then the assumption of typicality does not always give rise to the highest likelihoods. Understood from a Bayesian perspective, these results show that when one has the freedom to vary both the underlying theory and the xerographic distribution, one should find the combination of the two that maximizes the posterior probability, and then from this combination, one can \emph{infer} how typical we are. 

Through these three entwined aspects of typicality, therefore, I conclude (in section~\ref{SEC:Conclusion}) that the principle of mediocrity is more questionable than has been recently claimed (Gott 1993, Vilenkin 1995, Page 1996, Bostrom 2002, Garriga and Vilenkin 2008).

\section{{\bf\large{Typicality Matters}}}\label{SEC:TypMatters}

The problem of extracting predictions from multiverse scenarios is a difficult, almost forbidding, one. Indeed, the implicit optimism of the first line of section~\ref{SEC:Introduction} should be tempered with the warning that it is unclear that we presently \emph{have} a theory of the multiverse. What we do have are tentative extensions of more accepted theories into unfamiliar regimes of energy and time [such as the extension of inflation into the regime of eternal inflation---see Guth (2007) for a review], which hint at the existence of a multiverse. And what hints do exist paint a startling picture, set in the arena of a (presumably classical) spacetime, some parts of which undergo inflation \emph{forever}.

In this paper, we will set aside this warning, and turn instead to the question: assuming we do have such a theory of the multiverse, how will we be able to test it? We also set aside the issue of direct observation of the multiverse [although there are attempts to ascertain at least the local nature of the multiverse through collisions of other domains with ours---see Aguirre and Johnson (2011) for a review], and focus instead on how we can extract predictions for observables, such as parameters of the standard models of particle physics and cosmology, or other related parameters. Such predictions are naturally extracted from probability distributions over those parameters. Aguirre (2007) lays out seven related steps one needs to take in order to extract testable predictions. One can distill those steps into three main ones, which I now outline [see also Azhar and Butterfield (2016)]. This distillation is also useful because it maps our discussion onto the predictive scheme first outlined by Srednicki and Hartle (2010), which I will go on to endorse later in the paper. 

Assume then, that one has a theory $\mathcal{T}$ that describes a multiverse. There are three problems one needs to address, each of which brings one successively closer to a prediction based on a probability distribution generated from $\mathcal{T}$, and each of which is necessary to elicit a crisp prediction. 
\begin{itemize}
\item[(i)] \emph{The measure problem}: The first problem is the need  to specify the sample space, and in particular, the type of object $D$ that are elements of this sample space (to fix terminology: sets of these objects are events, over which probability measures are then defined). For example, a natural choice would be to fix $D$ to be a domain; but what if there are some domains that are much larger than others---should we count these larger domains in the same way as the smaller ones? The answers to such questions (which are not specified \emph{a priori}) dictate probabilities for various parameters defined over the $D$-objects. This problem is aggravated by the fact that often, measures over reasonable candidates for these objects are infinite. And though means of regularizing these infinities have been proposed, the resulting probabilities are sensitive to the regularization scheme employed. The combination of the lack of an unambiguous sample space, together with difficulties associated with infinities that arise in eternal inflation is called the `measure problem' [see, e.g., Freivogel (2011)]. 
\item[(ii)]\emph{The conditionalization problem}: It is likely that even with a solution to the measure problem in hand, probabilities for observables taking the values that we in fact measure will be small. So instead of then simply rejecting $\mathcal{T}$ (with its corresponding solution to the measure problem) on the grounds that it is not predictive, one \emph{conditionalizes} the probability distribution by: excising domains that do not satisfy specified criteria $\mathcal{C}$, and renormalizing the resulting probability distribution. But it is not clear what the criteria $\mathcal{C}$ should be; and different choices lead to different probabilities for observables, and thus, naturally, to different predictions (we develop examples of conditionalization schemes below). This problem is known as the `conditionalization problem' [see Aguirre and Tegmark (2005)]. 
\item[(iii)]\emph{The typicality problem}: Finally, even if one has determined a measure and a conditionalization scheme, one faces the issue, which we will largely focus on in this paper, of ascertaining how typical we are of the resulting domains. The assumption that we \emph{are} typical  is known as the \emph{principle of mediocrity}. Precisely what assumption we should adopt is the `typicality problem'. 
\end{itemize}

From a conceptual point of view, it is plausible that an appropriate measure and conditionalization scheme will render us more typical, and so the conditionalization scheme described in (ii) and the problem of typicality described in (iii) are correlated. But as I argued in Azhar (2014), inherent ambiguities in any chosen conditionalization scheme leave open a broad spectrum of plausible assumptions about typicality. 

For example, one could perform conditionalization in accordance with Carter's Weak Anthropic Principle (WAP) (Carter 1974; see also discussions in: Barrow and Tipler 1986, Earman 1987, and Bostrom 2002). This states: 
\begin{quote}
``\dots what we can expect to observe must be restricted by the conditions necessary for our presence as observers'' (Carter 1974, p.~291).
\end{quote}
So, one way forward would be to determine precisely what these conditions are---and an intimate part of this determination is figuring out exactly what is meant by `us'. Weinstein (2006) indeed points this out: that there is an ambiguity in the WAP amounting to whether it is referring to just `observers' or indeed more specifically, to us. In either case, I hold, there are problems associated with determining how to put the required constraint into concrete terms that we can input into our physical theories. And this problem remains if one tries to conditionalize only on (some precise version of) `our observational situation' (ostensibly avoiding mention of human/anthropic matters).

A plausible suggestion in light of these difficulties is to perform what is known as `top-down' conditionalization (Aguirre and Tegmark 2005). This is the idea that we conditionalize by fixing the values of all parameters in our theory to those that we have already observed, with a view towards then predicting the value of the observable we are actually interested in. There are, however, conceptual concerns with this approach [outlined in Azhar (2015, section V)]; such as the need to guarantee that the observable to be predicted is indeed correlated with the conditionalization scheme, but not overly so, as this would open one to the charge of circularity. Finding a balance in this situation is a difficult open problem. 

Setting aside conceptual or even technical difficulties in this situation, it is interesting to explore the question of whether in this highly constrained scenario, typicality makes a significant difference to predictions. For if, in a practical setting it doesn't, then one could use top-down conditionalization to significantly reduce the number of problems one faces in extracting predictions [as outlined in (i)--(iii) above]. 

Indeed, in Azhar (2014), I expressly address this issue, building on work by Aguirre and Tegmark (2005). The particular example I explore involves the prediction, in simplified multiverse settings, of the total number of species that contribute significantly to dark matter, under the assumption of top-down conditionalization.\footnote{Indeed, the possible existence of dark matter is well known, but it has not yet been ruled out that multiple distinct species contribute significantly to the total density of dark matter (see Bertone et al. 2005).} The conditionalization is achieved by fixing the total density $\rho$ (given by a dimensionless dark-matter-to-baryon ratio), to the observed value $\rho_{\textrm{\tiny obs}}$. This latter quantity was recently confirmed by the Planck Collaboration to be: $\rho_{\textrm{\tiny obs}}\approx 5$ (Ade et al. 2015). 

Aguirre and Tegmark (2005) show that under the assumption of typicality (i.e., where the prediction is determined by the peak of the relevant probability distribution over densities of dark matter species), and for probabilistically independent species, there will be multiple species that contribute significantly to the total dark matter density. In Azhar (2014), I go on to show that there are directions in the parameter space defined by densities of the various species, where the density of just a single dark matter species dominates, and that therefore, assumptions about typicality can change predictions in ways that are experimentally accessible. These results are further explored and supported under more general assumptions in Azhar (2016, section III): viz. where the densities of the various species can be probabilistically correlated or independent. 

Thus, in such (admittedly simplified) settings, typicality matters, both because it is not a superfluous conceptual addition to the problems discussed in (i) and (ii) above, and because even in the most constrained (top-down) scenarios, assumptions about typicality can significantly change physical predictions. 

\section{{\bf\large{Typicality and Problems of Confirmation}}}\label{SEC:Confirmation}

There is nothing in section~\ref{SEC:TypMatters} that suggests that the impact of typicality is lessened if we relax the constrained nature of the conditionalization scheme advocated there. Indeed, Aguirre and Tegmark (2005) discuss a `spectrum of conditionalization' that includes, in addition to top-down conditionalization: (i) the bottom-up approach, namely, no conditionalization of the theory-generated probability distribution at all; and (ii) `anthropic' conditionalization---which takes observers into account, in line with principles such as the WAP, but without being as stringent as top-down conditionalization. One would expect in each of these cases also, that typicality can play a significant role in impacting predictions. This is precisely what I explore in Azhar (2016); and in doing so, I uncover the existence of problems of confirmation, which I will discuss in this section. 

For bottom-up conditionalization, a unimodal theory-generated probability distribution $P(\vec{\rho}|\mathcal{T})$ is assumed, where $\vec{\rho}:=\{\rho_{i}\}_{i=1}^{N}$ corresponds to a collection of dimensionless dark-matter-to-baryon densities for some $N$ components of dark matter. It is shown that if we are given no further information (such as information about the location of the peak of the distribution), then the chance is small for the peak of the distribution to fall along the equal density diagonal in the $N$-dimensional space defined by the different species' densities. So, assuming that the range over which each of the species' densities can take values is much larger than $N$ [see Azhar (2016, section II) for a more precise version of this assumption], the expected number of species sharing the highest density, under the assumption of typicality, is just one [where, as in section~\ref{SEC:TypMatters}, typicality corresponds to being under the peak of a (conditionalized) theory-generated probability distribution]. Conversely, under atypicality, for a broad range of theory-generated distributions, this situation can change, and multiple species can contribute significantly to the total dark matter density. 

As mentioned in section~\ref{SEC:TypMatters}, the relationship between typicality and the number of species that contribute significantly to the total dark matter density can be different in the top-down case. For particular (simple) probability distributions, in Azhar (2016, sections III A and III B), I show that multiple species contribute significantly in the case of typicality, and that there exist directions in parameter space such that atypicality leads to a single dominant component. This is the opposite prediction to that found in the bottom-up case.\footnote{Of course, this conclusion depends on the assumed probability distribution, and in Azhar (2016, Section III C), I investigate one example in which typicality in the top-down approach corresponds to one species contributing significantly to the total dark matter density, and atypicality corresponds either to equal contributions from multiple species (viz. two species---in the restricted example studied there), or to a single species that is more dominant than in the case of typicality. More to the point, in more realistic cosmological settings, the relevant probability distribution should be uniquely determined by theoretical considerations (excepting ambiguities that may arise from the measure problem), and thus, for top-down conditionalization (or indeed any other conditionalization scheme), just one broad set of conclusions resulting from considerations of typicality will arise.}

The anthropic case is more subtle. The question of how one should conditionalize in accordance with anthropic considerations is, as I mentioned in section~\ref{SEC:Introduction}, fraught with conceptual and technical difficulties. To make progress on this issue, Aguirre and Tegmark (2005) propose that a weighting factor $W(\rho)$ can be introduced, that multiplies the theory-generated probability distribution $P(\vec{\rho}|\mathcal{T})$, and which expresses the probability of finding domains in which \emph{we} might exist as a function of (in this case) the total density of dark matter $\rho:=\sum_{i=1}^{N}\rho_{i}$. The product of $P(\vec{\rho}|\mathcal{T})$ and $W(\rho)$ is then renormalized to give a distribution that putatively takes anthropic considerations into account (in this admittedly simplified setting). Of course, how faithfully this approach relates to `observers', is tied to the degree to which $W(\rho)$ faithfully represents probabilities for observers as a function of the parameter of interest (i.e., $\rho$). In the absence of concrete means to determine this connection, we can only speculate about what sorts of predictions such a scheme might yield. 

And so in Azhar (2016), I take up such speculations by assuming a particular form for $W(\rho)$ whose main feature is a Gaussian fall-off as a function of the total dark matter density $\rho$ [as in Aguirre and Tegmark (2005)]. I then explore how assumptions about typicality can change the prediction for the total number of species $N_{\tiny\textrm{EQ}}$, that contribute equally to the total dark matter density ({\small `EQ'} for `equal'). That is, a variation on the nature of the discussion presented in the bottom-up and top-down cases is explored for this conditionalization scheme, where one can bound the total number of species $N$ itself (in the other two cases, $N$ is fixed by assumption at the outset). I find that atypicality can indeed affect predictions for $N_{\tiny\textrm{EQ}}$ in significant ways. As just one example of this effect [see figure 3a in Azhar (2016)], the assumption of typicality for correlated species of dark matter that are distributed in a Gaussian manner, can yield a total of $N_{\tiny\textrm{EQ}}= 2\;\textrm{or}\;5$, whereas under atypicality, $N_{\tiny\textrm{EQ}}= 7$. One thus finds that atypicality has a measurable effect on the total number of equally contributing species of dark matter. 

For each of the three conditionalization schemes discussed above, it is clear that typicality matters; and it turns out that it does so in such a way as to lead to coincident predictions. In the examples above, these predictions were for the total number of significantly contributing species of dark matter, as a function of what Srednicki and Hartle (2010) call a `framework': namely, the conjunction of a theory that gives rise to a multiverse (including an associated measure), a conditionalization scheme, and an assumption about typicality. Indeed as shown in Azhar (2016), different frameworks can generate the same prediction for the total number of species that contribute significantly to the total dark matter density---whether that number comprises just one dominant species or multiple species. 

From the point of view of confirmation, this insight implies that the experimental determination of the total number of species that contribute significantly to dark matter would not distinguish between frameworks. That is, there is a significant under-determination of multiverse frameworks by data. In more realistic cosmological circumstances then, if such under-determination is robust to the choice of observables we aim to predict the values of, and we do not invoke more intricate confirmation schemes---such as Bayesian analysis, which would allow us to invoke priors over frameworks to help in their demarcation---then we must conclude that our observations alone won't be enough to confirm a single multiverse framework. 

\section{{\bf\large{Typicality as the Best Strategy}}}\label{SEC:testingTypicality}

Thus far I have argued that assumptions about typicality are important, since they can make a difference to predictions derived from theories of the multiverse (sections~\ref{SEC:TypMatters} and~\ref{SEC:Confirmation}). And different typicality assumptions, when included as part of various multiverse frameworks, can lead to the same prediction (section~\ref{SEC:Confirmation}). So the question arises: is there an independent means by which we can favour certain typicality assumptions, so as to guide predictions generated from multiverse theories, as well as to aid in the task of framework confirmation?

Srednicki and Hartle (2010) develop a Bayesian scheme that helps in addressing this issue. As mentioned in section~\ref{SEC:Confirmation}, they contend that a prediction arises from a framework: namely, the \emph{conjunction} of a theory $\mathcal{T}$ that describes a multiverse (or a ``very large universe'', in their language), a conditionalization scheme $\mathcal{C}$, and an assumption about typicality that for now, we will label abstractly by $\xi$. Thus a prediction for data that we observe today, denoted by $D_{0}$, is made via a theory-generated probability distribution $P(D_{0}|\mathcal{T}, \mathcal{C}, \xi)$, that is properly considered to be conditional on some framework $\{\mathcal{T}, \mathcal{C}, \xi\}$. If the prediction we extract from the framework is not verified in experiments, we have licence to change any one of its conjuncts, and to then reassess the predictive power of the new framework. Indeed, when we have a variety of frameworks at our disposal, we can formalize the task of confirmation through Bayesian analysis---and this is  what Srednicki and Hartle endorse. That is, one constructs the posterior distribution $P(\mathcal{T}, \mathcal{C}, \xi|D_{0})\propto P(D_{0}|\mathcal{T}, \mathcal{C}, \xi)P(\mathcal{T}, \mathcal{C}, \xi)$, where $P(\mathcal{T}, \mathcal{C}, \xi)$ is a prior over the framework given by $\{\mathcal{T}, \mathcal{C}, \xi\}$, and then the framework with the highest posterior probability is the one that is confirmed. 

Typicality assumptions, i.e., various $\xi$'s, are introduced via what Srednicki and Hartle dub `xerographic distributions'. These are probability distributions that express, by \emph{assumption}, the probability that we are at some location in a multiverse where a member of our reference class exists (generally, our reference class will have many  members at different locations throughout the multiverse). To be more precise, xerographic distributions are most easily understood in the situation where the conditionalization scheme used to distinguish locations of interest yields a finite number of locations. If we assume that we are typical members of our reference class, the probability that we are at any one such location is the same, so that the xerographic distribution is the uniform distribution. Assumptions of atypicality can be expressed via non-uniform distributions over these locations. 

It is broadly within this schema that in Azhar (2015) I look to `test' the principle of mediocrity. I do this for a simple multiverse model that extends a model first introduced by Hartle and Srednicki (2007). The model I consider has a total of $N$ domains in the multiverse. Each domain either has observers in it or it does not (so the existence of observers can be mapped onto a binary variable in each domain---the total number of observers in any domain is not considered). There exists a single observable that can take one of two possible values (distinguished simply by colour---either red or blue). The assumption is made that we know (i) we exist (i.e., there exists at least one domain with observers in it) and (ii) we see a particular value of the observable (i.e., we see red).

The frameworks I consider consist of a set of theories $\mathcal{T}$ that specify just the colour of each domain. The conditionalization scheme implemented is, in effect, a limiting case of top-down conditionalization. Though, to bring both notation and terminology in the remainder of this section in line with Azhar (2015), I will drop explicit mention of this conditionalization scheme, and will refer to frameworks as consisting of a theory and a xerographic distribution defined over some reference class. The uniform distribution that implements the principle of mediocrity is defined against a maximally specific reference class: namely, observers who see our data [i.e., observers who see red; note also that individual `locations' over which xerographic distributions can be defined are just individual domains in this case]. Atypicality is then implemented by assuming uniform distributions over \emph{different} reference classes (for example, the reference class of observers, regardless of the colour they see). 

These considerations allow us to compute likelihoods for our data $D_{0}:=$ \emph{there exist observers who see red}. These likelihoods can be written, in the above notation, as $P(D_{0}|\mathcal{T}, \xi)$, where again, $\xi$ represents some xerographic distribution, with $\xi^{\tiny\textrm{PM}}$ representing the xerographic distribution that implements the principle of mediocrity. I assume that each of the frameworks that enter into the analysis do so with an equal prior, so that the posterior probability of a framework is proportional to its likelihood: $P(\mathcal{T}, \xi|D_{0})\propto P(D_{0}|\mathcal{T}, \xi)$. `Testing typicality' then amounts to comparing appropriate likelihoods against one another: namely, comparing the relative sizes of $P(D_{0}|\mathcal{T}, \xi^{\tiny\textrm{PM}})$ and $P(D_{0}|\mathcal{T}^{\star},  \xi^{\star})$; where $\xi^{\star}\neq  \xi^{\tiny\textrm{PM}}$ and where the theory $\mathcal{T}$ is allowed to vary as well. 

I find two main results. The first is the fact that for a fixed theory, the xerographic distribution that implements the principle of mediocrity gives rise to the highest likelihoods for our data---so that by one measure, the principle of mediocrity does well. However (and this is the second main result), it is not universally the case that the principle of mediocrity provides the highest likelihoods. This second finding amounts to the statement that although for some theory $\mathcal{T}_{1}$, the xerographic distribution that implements the principle of mediocrity, $\xi^{\tiny\textrm{PM}}$, provides the highest likelihoods for our data, it may well be that there is a second theory $\mathcal{T}_{2}$, which, when partnered with a xerographic distribution that does not implement the principle of mediocrity, gives rise to a \emph{higher} likelihood than the framework $\{\mathcal{T}_{1},\xi^{\tiny\textrm{PM}}\}$. This is particularly pertinent when $\mathcal{T}_{2}$ is a theory for which the principle of mediocrity is not a viable partner. 

For example, there exist theories in which `Boltzmann brains' exist and outnumber ordinary observers [see Albrecht and Sorbo~(2004), and De Simone et al.~(2010)]. Boltzmann brains can observe our data $D_{0}$, but generally, their experimental record is disordered and uncorrelated with the past. Under such circumstances, it may well be that likelihoods for our data are highest when we presume we are typical observers in this scenario, but an unwanted consequence of this assumption is that we are then likely to \emph{be} Boltzmann brains---namely, the theory, in conjunction with this typicality assumption (and the conditionalization scheme that is implicit throughout this discussion) is likely to predict that our future experiments will be disordered and uncorrelated with our past experiments. In such a situation, we would like to discount the principle of mediocrity, viz. $\xi^{\tiny\textrm{PM}}$, as a plausible partner of this theory, without necessarily discarding the theory that gives rise to Boltzmann brains in the first place. Precisely this freedom is allowed by the scheme (described above) developed by Srednicki and Hartle (2010).

Now, the question arises: why would we want to support a theory in which we are not typical? Hartle and Srednicki (2007) provide an example that exposes a fallacy in this line of questioning: we are simply not justified in discarding a theory just because we would not be typical according to the theory. For example, if some theory predicted the existence of many more sentient beings on Jupiter than on Earth, then, \emph{ceteris paribus}, it is unjustified to discard the theory just because we are atypical according to it. Combining the lesson from this example and the second main result obtained in Azhar (2015), I conclude that it is legitimate to consider a framework that describes us as atypical, and it may well be that this framework is better confirmed in Bayesian terms than a framework (with a different theory) that implements the principle of mediocrity.

\section{{\bf\large{Conclusion}}}\label{SEC:Conclusion}

The principle of mediocrity is thus rather controversial in multiverse cosmology. It stems, of course, from our intuitions about how to reason in more controlled settings (such as in laboratory experiments), but its application to scenarios that we are less familiar with, such as those presented by cosmological theories that describe a multiverse, is fraught with complications. 

As we have surveyed above, these complications involve (at least): (i) the problem that we cannot simply assume typicality, as it is far from clear who or what we restrict attention to when we conditionalize theory-generated probability distributions (even in top-down scenarios)---indeed, this is a live issue because assumptions about typicality can change the prediction for what we should expect to see; (ii) the issue that typicality matters in such a way as to lead to a severe under-determination of multiverse frameworks by data; and (iii) that the principle of mediocrity may not be the assumption that is the most predictive of our data, or indeed the most confirmed, in Bayesian terms. 

It remains to ascertain the extent to which the above claims [(i)--(iii)], developed in rather stylized settings, apply to more realistic cosmological scenarios [see, e.g., Hartle and Hertog (2013, 2015, 2016)]. But for now, we must maintain that the principle of mediocrity is more questionable than has been generally claimed. 
\\

{\em Acknowledgements}:--- I thank Jeremy Butterfield, Hasok Chang, Jim Hartle, and Dean Rickles for discussions, as well as audiences in Sydney, Cambridge, M\"{u}nchen, and D\"{u}sseldorf. I am supported by the Wittgenstein Studentship in Philosophy at Trinity College, Cambridge. 
\vspace{0.275cm}
\section*{{\bf\large{References}}}\label{SEC:References}

\begin{hangparas}{.25in}{1} 
Ade, P. A. R., et al.~(\emph{Planck} Collaboration). 2015. \emph{Planck} 2015 results. XX. Constraints on inflation. 	arXiv:1502.02114 [astro-ph.CO]. \url{https://arxiv.org/abs/1502.02114}.\\

Aguirre, Anthony. 2007. Making predictions in a multiverse: Conundrums, dangers, coincidences. In \emph{Universe or multiverse?}, ed. Bernard Carr, 367--386. Cambridge: Cambridge University Press. arXiv:astro-ph/0506519.\\

Aguirre, Anthony, and Matthew C. Johnson. 2011. A status report on the observability of cosmic bubble collisions. \emph{Reports on Progress in Physics} 74: 074901. arXiv:0908.4105 [hep-th].\\

Aguirre, Anthony, and Max Tegmark. 2005. Multiple universes, cosmic coincidences, and other dark matters. \emph{Journal of Cosmology and Astroparticle Physics} 01(2005)003. arXiv:hep-th/0409072.\\

Albrecht, Andreas, and Lorenzo Sorbo. 2004. Can the universe afford inflation?, \emph{Physical Review} D 70: 063528. arXiv:hep-th/0405270.\\ 

Azhar, Feraz. 2014. Prediction and typicality in multiverse cosmology. \emph{Classical and Quantum Gravity} 31: 035005. arXiv:1506.08101 [astro-ph.CO].\\

Azhar, Feraz. 2015. Testing typicality in multiverse cosmology. \emph{Physical Review} D 91: 103534. arXiv:1506.05308 [physics.hist-ph].\\

Azhar, Feraz. 2016. Spectra of conditionalization and typicality in the multiverse. \emph{Physical Review} D 93: 043506. arXiv:1601.05938 [astro-ph.CO].\\

Azhar, Feraz, and Jeremy Butterfield. 2016. Scientific realism and primordial cosmology. Invited contribution for \emph{The Routledge handbook on scientific realism}. ed. Juha Saatsi. London: Routledge. arXiv:1606.04071 [physics.hist-ph]. \url{https://arxiv.org/abs/1606.04071}. PhilSci archive: \url{http://philsci-archive.pitt.edu/12192/}.\\

Barrow, John D., and Frank J. Tipler. 1986. \emph{The anthropic cosmological principle}. Oxford: Oxford University Press.\\

Bertone, Gianfranco, Dan Hooper, and Joseph Silk. 2005. Particle dark matter: Evidence, candidates and constraints. {\it Physics Reports} 405: 279--390. arXiv:hep-ph/0404175.\\

Bostrom, Nick. 2002. \emph{Anthropic bias: Observation selection effects in science and philosophy}. New York: Routledge.\\

Bousso, Raphael, and Joseph Polchinski. 2000. Quantization of four-form fluxes and dynamical neutralization of the cosmological constant. \emph{Journal of High Energy Physics} JHEP06(2000)006. arXiv:hep-th/0004134.\\ \\

Carter, Brandon. 1974. Large number coincidences and the anthropic principle in cosmology. In \emph{Confrontation of cosmological theories with observational data}. IAU Symposium No. 63, ed. M. S. Longair, 291--298. Dordrecht: Reidel.\\

De Simone, Andrea, Alan H. Guth, Andrei Linde, Mahdiyar Noorbala, Michael P. Salem, and Alexander Vilenkin.  2010. Boltzmann brains and the scale-factor cutoff measure of the multiverse. {\em Physical Review} D 82: 063520. arXiv:0808.3778 [hep-th].\\

Earman, John. 1987. The SAP also rises: A critical examination of the anthropic principle. \emph{American Philosophical Quarterly} 24: 307--317.\\

Freivogel, Ben. 2011. Making predictions in the multiverse. \emph{Classical and Quantum Gravity} 28: 204007. arXiv:1105.0244 [hep-th].\\

Garriga, Jaume, and Alexander Vilenkin. 2008. Prediction and explanation in the multiverse. \emph{Physical Review} D 77: 043526. arXiv:0711.2559 [hep-th].\\

Gott, J. Richard., III. 1993. Implications of the Copernican principle for our future prospects. \emph{Nature} 363: 315--319.\\

Guth, Alan H. 2007. Eternal inflation and its implications. \emph{Journal of Physics A: Mathematical and Theoretical} 40: 6811--6826. arXiv:hep-th/0702178.\\

Hartle, James B. 2007. Anthropic reasoning and quantum cosmology. In \emph{Universe or multiverse?}, ed. Bernard Carr, 275--284. Cambridge: Cambridge University Press.\\

Hartle, James, and Thomas Hertog. 2013. Anthropic bounds on $\Lambda$ from the no-boundary quantum state. \emph{Physical Review} D 88: 123516. arXiv:1309.0493 [astro-ph.CO].\\

Hartle, James, and Thomas Hertog.~2015. The observer strikes back.~arXiv:1503.07205 [gr-qc]. \url{https://arxiv.org/abs/1503.07205}.\\

Hartle, James, and Thomas Hertog. 2016. One bubble to rule them all. arXiv: 1604.03580 [hep-th]. \url{https://arxiv.org/abs/1604.03580}.\\

Hartle, James B., and Mark Srednicki. 2007. Are we typical? \emph{Physical Review} D 75: 123523. arXiv:0704.2630 [hep-th].\\

Kachru, Shamit, Renata Kallosh, Andrei Linde, and Sandip P. Trivedi. 2003. de Sitter vacua in string theory. \emph{Physical Review} D 68: 046005. arXiv:hep-th/0301240.\\

Linde, Andrei D. 1983. The new inflationary universe scenario. In \emph{The Very Early Universe, Proceedings of the Nuffield Workshop, Cambridge, 21 June to 9 July, 1982}, eds. G. W. Gibbons, S. W. Hawking and S. T. C. Siklos, 205--249. Cambridge: Cambridge University Press.\\

Linde, Andrei D. 1986a. Eternal chaotic inflation. \emph{Modern Physics Letters} A 01: 81--85.\\

Linde, Andrei D. 1986b. Eternally existing self-reproducing chaotic inflationary universe. \emph{Physics Letters} B 175: 395--400.\\

Page, Don N. 1996. Sensible quantum mechanics: Are probabilities only in the mind? \emph{International Journal of Modern Physics} D 05: 583--596. arXiv:gr-qc/9507024.\\

Srednicki, Mark, and James Hartle. 2010. Science in a very large universe. \emph{Physical Review} D 81: 123524. arXiv:0906.0042 [hep-th].\\

Steinhardt, Paul J. 1983. Natural inflation. In \emph{The Very Early Universe, Proceedings of the Nuffield Workshop, Cambridge, 21 June to 9 July, 1982}, eds. G. W. Gibbons, S. W. Hawking and S. T. C. Siklos, 251--266. Cambridge: Cambridge University Press.\\

Susskind, Leonard. 2003. The anthropic landscape of string theory. arXiv:hep-th/0302219. \url{https://arxiv.org/abs/hep-th/0302219}. Published in \emph{Universe or multiverse?}, ed. Bernard Carr, 247--266, (2007). Cambridge: Cambridge University Press.\\

Vilenkin, Alexander. 1983. Birth of inflationary universes. \emph{Physical Review} D 27: 2848--2855.\\

Vilenkin, Alexander. 1995. Predictions from quantum cosmology. \emph{Physical Review Letters} 74: 846--849. arXiv:gr-qc/9406010.\\

Weinstein, Steven. 2006. Anthropic reasoning and typicality in multiverse cosmology and string theory. \emph{Classical and Quantum Gravity} 23: 4231--4236. arXiv:hep-th/0508006.

\end{hangparas}
\end{document}